\documentclass[aps,prev,twocolumn,preprintnumbers,floatfix,nofootinbib]{revtex4-1}
\pdfoutput=1
\usepackage{graphicx}
\usepackage{bm}
\usepackage{times}
\usepackage{slashed}
\usepackage{color}
\usepackage{color}

\newcommand{\be}{\begin{equation}}
\newcommand{\ee}{\end{equation}}
\newcommand{\bea}{\begin{eqnarray}}
\newcommand{\eea}{\end{eqnarray}}

\begin{document}

\title{The Poker Face of the Majoron Dark Matter Model: LUX to keV Line}
%\title{Is the Majoron a Viable Dark Matter Candidate?}
%\title{A Simplified Model For Letogenesis, See-Saw and Majoron Dark Matter}

\author{Farinaldo S. Queiroz}
\email{fdasilva@ucsc.edu}

\author{Kuver Sinha}
\email{kusinha@syr.edu}

\affiliation{Department of Physics and Santa Cruz Institute for Particle
Physics University of California, Santa Cruz, CA 95064, USA\\
Department of Physics, Syracuse University, Syracuse, NY 13244, USA}

\begin{abstract}
We study the viability of pseudo Nambu-Goldstone bosons (Majorons) arising in see-saw models as dark matter candidates. Interestingly the stability of the Majoron as dark matter is related to the scale that sets the see-saw and leptogenesis mechanisms, while its annihilation and scattering cross section off nuclei can be set through the Higgs portal. For $\mathcal{O}(GeV) - \mathcal{O}(TeV)$ Majorons, we compute observables such as the abundance, scattering cross section, Higgs invisible decay width, and emission lines and compare with current data in order to outline the excluded versus still viable parameter space regions. We conclude that the simplest Majoron dark matter models coupling through the Higgs portal, except at the Higgs resonance, are excluded by current direct detection data for Majorons lighter than $225$~GeV and future runnings are expected to rule out decisively the 1GeV-1TeV window. Lastly, we point out that light keV-scale Majorons whose relic density is set by thermal freeze-in from sterile neutrinos can account for the keV line observed by XMM-Newton observatory in the spectrum of 73 galaxy clusters, within a see-saw model with a triplet Higgs. 

\end{abstract}

\pacs{95.35.+d, 14.60.Pq, 98.80.Cq, 12.60.Fr}

\maketitle

\section{Introduction}

The identity of dark matter constitutes one of the most exciting puzzles in current science. 
%There is a huge variety of dark matter candidates in the literature with the so called WIMPs (Weakly Interacting Massive Particles) being the most studied ones. In principle any particle physics model is capable of having a WIMP as dark matter as long as it has a $SU(2)_L$ singlet neutral particle which is stabilized by some sort of discrete or global symmetry. 
Interestingly, dark matter is often connected to other paradigms of fundamental physics, with WIMPs (Weakly Interacting Massive Particles) in supersymmetric theories being the most studied ones. The dark sector can also be connected to other important phenomena such as the generation of the neutrino masses, leptogenesis, and baryogenesis \cite{Tsuyuki:2014aia,Kashiwase:2013uy,seesaw}. The Majoron dark matter model is an example which occurs in see-saw models of neutrino mass generation. 

%A popular method of generating neutrino masses is the see-saw mechanism, which also naturally accommodates leptogenesis. 
In see-saw models, the lepton number might be explicitly broken by the Majorana masses of right-handed neutrinos. If the lepton number is instead broken spontaneously by the vev of a complex scalar field (a singlet ``Higgs"), one has a new pseudo-scalar gauge singlet Nambu-Goldstone boson (the Majoron). In such models, the Majoron is a natural decaying dark matter candidate \cite{ernestma,CMBmajoron,Gelmini:1984pe,Berezinsky:1993fm}. The Majoron lifetime is determined by its decay into Standard Model (SM) neutrinos which is suppressed by the scale of lepton number violation. For sufficiently high lepton violation scale the Majoron is cosmologically stable. This scale also sets the heavy right handed neutrinos masses in the see-saw type I setup. Therefore, the viability of the Majoron dark matter is connected to the see-saw mechanism responsible for generating the SM neutrinos masses and the scale of leptogenesis which occurs through the decays of the heavy right-handed neutrinos \cite{Covi:2009xn}.  

As for the mass of the Majoron, it can arise from explicit soft terms, or from quantum gravitational effects that explicitly break lepton number. In the former case, the mass can be hundreds of GeV. From an effective field theory point of view, nothing prevents a coupling of the Majoron to the Higgs scalar potential at tree level. Majoron models of this category are thus a particular UV realization of the effective Higgs-portal scalar models studied by Ref.\cite{Burgess:2000yq}. In the case where the mass is due to quantum gravitational effects, on the other hand, the Majoron is expected to be very light. A particularly well-motivated scenario is a $\mathcal{O}($keV$)$ Majoron, which can satisfy the thermal relic density.

We study dark matter observables in both cases. In the former,  we compute %direct and indirect %Majoron 
observables such as the abundance, scattering cross section, Higgs invisible decay width, and emission lines and compare with current data in order to outline the excluded versus still viable parameter space regions in the $\mathcal{O}(GeV) - \mathcal{O}(TeV)$ mass range. We find that LUX bounds on the dark matter scattering cross section, along with relic density requirements and Higgs invisible decay width limit, effectively rule out thermal dark matter below $\sim 225$ GeV in the this scenario (except near the Higgs resonance when the Majoron mass is $\sim 60$ GeV). Furthermore, future direct detection running coming from XENON1T and LUX are forecasted to rule out the entire GeV-TeV mass range. A way out would be the inclusion of new particles with masses close to the Majoron in order to exploit co-annihilation channels. Additionally, a non-thermal history for the Majoron, where the initial number density of Majorons is fixed without further annihilation, is possible \cite{Allahverdi:2010rh}.

In the case of a $\mathcal{O}(keV)$ Majoron, we point out that it may be possible to accommodate the recently observed keV line from the XMM-Newton observatory \cite{KeVline}, for appropriate choice of parameters in a see-saw model with a triplet (as well as the singlet) Higgs scalar. The branching to photons arises at loop level from the projection of the Majoron along the doublet (Standard Model) Higgs. The relic density can be set by thermal freeze-in from the right-handed neutrinos. We provide  order-of-magnitude estimates as a proof of concept that the observed keV line can be obtained in this class of models.

The plan of the paper is as follows. In Section \ref{model}, we describe the Majoron dark matter model in the context of type I see-saw. In Section \ref{darkmatter}, we discuss the dark matter observables of the model. In Section \ref{kevmajoron}, we discuss the case of a light Majoron and the recently observed keV line. We end with our Conclusions.

\section{Model} \label{model}

We will study a  model in which the leptogenesis conditions, the see-saw mechanism and existence of a dark matter candidate are connected. The model is comprised of a neutral singlet scalar and three right-handed neutrinos. Therefore the Lagrangian of the model
might be written as,
\begin{eqnarray} \label{yukawa}
\mathcal{L} \supset - \lambda \bar{L}^{}\Phi
N_{R}-\frac{1}{2}h \bar{N}_{R}^{c}N_{R} \sigma +\textrm{h.c.}\,
\end{eqnarray}where $L$ and $\Phi$ are the SM standard model lepton and Higgs doublets respectively, whereas $N_R$ are the right-handed neutrinos and
$\sigma$ is the singlet scalar. According to Eq.\ref{yukawa} we notice that the right-handed neutrinos carry a lepton number of 1, as well as the SM particles, whereas the neutral scalar carries a lepton number of $L=-2$.
The general scalar potential of the fields is found to be, 
\begin{eqnarray}
\label{potential}
V(\xi,\phi)&=&-\mu_{1}^{2}\sigma^{\dagger}_{}\sigma+\lambda_{1}(\sigma^{\dagger}_{}\sigma)^{2}_{}-\mu_{2}^{2}\Phi^{\dagger}_{}\Phi
+\lambda_{2}^{}(\Phi^{\dagger}_{}\Phi)^{2}_{}\nonumber\\
&+&2\lambda_3^{}\sigma^{\dagger}_{}\sigma\Phi^{\dagger}_{}\Phi + V_{{\rm soft}}\,,
\end{eqnarray}where $V_{{\rm soft}}$ is the term that softly breaks the global lepton number given by,
\begin{eqnarray}
\label{potential2}
V_{soft}&=&-\frac{1}{2}\mu_{3}^{2}(\sigma^{2}_{}+\textrm{h.c.})\,.
\end{eqnarray}
In the scenario where $V_{soft} \equiv 0$, the spontaneous symmetry breaking of the global lepton number, caused by the vev of the real component of the neutral scalar $\sigma$, will induce a massless Majoron in the theory. Conversely when $V_{soft} \neq 0$ is as defined above, the Majoron will acquire mass proportional to $\mu_{3}$ as we will see further. Additional soft terms such as the presence of Majorana mass terms for the right-handed neutrino may be evoked in more general settings. In this work we will assume the existence of a $Z_4$ discrete symmetry with the fields transforming as: $\phi \rightarrow \phi, \sigma \rightarrow -\sigma$ and $f \rightarrow -if$, where f stands for all fermions including the right-handed neutrinos with the remaining fields transforming trivially under $Z_4$. Notice that the Lagrangians aforementioned are invariant under this symmetry. Throughout our study we will keep $\lambda_{1,2} >0$ and $\lambda_3 > -\sqrt{\lambda_1 \lambda_2}$ to guarantee the potential bounded from below. Since $\sigma$ is a complex scalar, we can write it as,
\begin{eqnarray}
\sigma=\frac{1}{\sqrt{2}}(\frac{u +\chi}{\sqrt{2}} +i J)\,.
\end{eqnarray}
Here u refers to the scale at which the lepton number is violated. In this work we assume u to lie at the GUT scale for leptogenesis and dark matter purposes. Anyway, after the spontaneous symmetry breaking mechanism, we find \cite{ernestma}
\begin{eqnarray}
\label{mass} \mathcal{L}_{m}^{}\supset
-\frac{1}{2}(\chi,H,J)\left[
\begin{array}{ccc}
2\lambda_1^{}u^2_{}&2\lambda_3^{}uv&0\\
[3mm] 2\lambda_3^{}uv&2\lambda_2^{}v^2_{}&0\\
[3mm] 0&0&2\mu_3^2 \end{array} \right]\left[
\begin{array}{c}
\chi\\
[3mm] H\\
[3mm] J\end{array} \right]\,,
\end{eqnarray}where we have used the standard definition for the Higgs doublet with $\Phi^T=1/\sqrt{2}\left[v+H,0\right]$ and identified the following relations,
\begin{eqnarray}
u&=&\sqrt{2}\langle\sigma\rangle =\sqrt{
\frac{\lambda_2^{}(\mu_1^2+\mu_3^2)-\lambda_3^{}\mu_2^2}{\lambda_1^{}\lambda_2^{}-\lambda_3^2}}\,,\\
w&=&\sqrt{2}\langle\chi\rangle =0\,,\\
v&=&\sqrt{2}\langle\phi\rangle =\sqrt{
\frac{\lambda_1^{}\mu_2^2-\lambda_3^{}(\mu_1^2+\mu_3^2)}{\lambda_1^{}\lambda_2^{}-\lambda_3^2}}\,.
\end{eqnarray}
From Eq.\ref{mass} we conclude the Majoron (J) is decoupled from the other scalars of the model. Additionally, due to the high hierarchy between assumed GUT scale vev u the electroweak vev (v), the mass matrix in Eq.\ref{mass} yields \cite{ernestma},
\begin{eqnarray}
m_{\chi}^{2}\simeq 2\lambda_1 u^2_{}\,,~~ m_{H}^{2}\simeq
2\lambda_2 v^2_{}\,,~~
m_J^2=2\mu_3^2\,.
\end{eqnarray}
\subsection{See-saw Type I}
See-saw type I also known as canonical see-saw refers to the mechanism where
the standard model neutrinos acquire masses through the insertion of the heavy
right-handed neutrinos as in our model. After the spontaneous symmetry breaking Eq.(\ref{yukawa}) turns into,
\begin{equation}
\mathcal{L} \supset \overline{\nu_L} M_D N_R + \frac{1}{2} \overline{N^C_L}M_R N_R +h.c.,
\label{Nreq}
\end{equation}which can be rewritten as 
\begin{eqnarray}
\mathcal{L} &=& \frac{1}{2}\left(\begin{array}{cc}\overline{\nu_L} &
    \overline{ {N}^C_L}\end{array}\right)\left(\begin{array}{cc} 0 &
      M_D \\ M_D^T & M_R\end{array}\right)\left(\begin{array}{c}{\nu}^C_R\\
     N_R\end{array}\right)+h.c.,
\label{eq-07:lagml}
\end{eqnarray}and yields,
\begin{equation}
m_\nu = M^T_D M^{-1}_R M_D.
\label{seesawI}
\end{equation}where according to Eq.(\ref{yukawa}) we find,
\begin{eqnarray}  
M_D & = &\frac{1}{\sqrt{2}}\lambda v \nonumber\\
M_R & = &\frac{1}{\sqrt{2}}hu\,.
\end{eqnarray}
For $M_R<10^{14}\,\textrm{GeV}$ the standard model neutrino masses can lie at the eV range naturally small with $m_D^{}$ being at the weak scale. The leptogenesis mechanism is not the focus in this work, however it is important to point out that the presence of heavy right-handed neutrinos coupled to the left-handed SM neutrinos through the interactions given in the first term of Eq.\ref{Nreq} offers an important connection to leptogenesis because the decay of the right-handed neutrinos and their annihilations into SM particles can generate a lepton asymmetry in non-CP conserving setups \cite{Gu:2009hn}. For $M_{N_R} > 10^9$~GeV and $u \sim 10^{15}$~GeV, the desired CP asymmetry is induced as long as the annihilations of the right-handed neutrinos into the massless Majorons go out of equilibrium before the sphaleron process is over \cite{Gu:2009hn}. 

Now  we will turn our focus on the dark matter observables.

\section{Abundance, Direct and Indirect Detection and Higgs Bounds} \label{darkmatter}

Viable dark matter candidates are either truly stable or cosmologically stable in the sense that their lifetime is much bigger than the age of the universe. In our model the Majoron belongs to the latter case as we will explain further. After the spontaneous symmetry breaking Eq.(\ref{yukawa}) turns into, 

\begin{eqnarray}
\label{yukawa3} \mathcal{L}\supset
-\frac{i}{2\sqrt{2}}h J \bar{N}_R^{c}N_R^{}+\textrm{H.c.}\,.
\end{eqnarray}where J is the pseudo-Majoron and the dark matter candidate of the model. We can clearly see that in the mass regime of interest that is $M_{J}\ll M_R$, the Majoron will decay into SM particles via virtual right-handed neutrinos which are expected to be super heavy due to the large value of the vev {\it u}. The decay width is dominated by the two-body decay into SM neutrinos according to, %\cite{ernestma,CMBmajoron},
\begin{eqnarray}
\Gamma_{J}^{} =\frac{1}{16 \pi}
\frac{\Sigma_i^{} m_{\nu_i}^2}{u^{2}_{}}m_{J}\,,
\label{Eqdecaywidth}
\end{eqnarray}which gives 
\begin{eqnarray}
\tau_{J} & = &\left(\frac{0.01\,\textrm{eV}^2_{}}{\Sigma_i^{} m_{\nu_i^{}}^2}\right)\left(\frac{u}{
5.5\times10^{15}_{}\,\textrm{GeV}}\right)^2_{}\left(\frac{\textrm{1\,TeV}}{m_J}\right)\nonumber\\
&&\times 10^{26}_{}\,\textrm{sec}\,.
\label{Eqlifetime}
\end{eqnarray}
Hence for the mass range of interest a singlet Majoron is cosmologically stable and in principle a viable dark matter candidate. In our model the Majoron is thermally produced via the Higgs portal according to the interactions,
\begin{eqnarray}
V\supset\lambda_3 J ^2 \Phi^\dagger\Phi\Rightarrow
\lambda_3^{} v J^2 H+\frac{1}{2}\lambda_3^{}
J^2 H^2\,.
\label{jjhh}
\end{eqnarray}
Therefore the parameter $\lambda_3$ play a crucial role in this model because it connects
the dark and visible sector of the model. Moreover, this coupling between pseudo-Majoron $J$ to the SM Higgs $H$ sets the relic abundance as well as the scattering cross section. At the end of the day we are left with two free parameters only which are $\lambda_3$ and the Majoron mass. Note for a sizeable $\lambda_3$, the present symmetry breaking pattern, i.e. $u \sim 10^{16}_{}\,\textrm{GeV}\gg v\simeq 246\,\textrm{GeV}$,
requires a large cancellation between $\lambda_3^{}u^2_{}$ and $\mu_2^2$ so that $\lambda_3^{}u^2_{}-\mu_2^2$ can be of the order of $v^2$ \cite{ernestma}. 

First we would like to derive Higgs bounds on 
the Majoron model. For a Majoron lighter than half of the
Higgs mass, the h $\rightarrow JJ$ is 
kinematically available and therefore it alters the measured
invisible width of Higgs. The current limit on the
branching ration into invisible particles is around
$10-15\%$ \cite{Belanger:2013xza,Ellis:2013lra}. A projected bound of 
$5\%$ at $14$~TeV LHC after $300 {\rm fb^{-1}}$ has been claimed \cite{Peskin:2012we}.
We use the latter one as reference with
no impact on our conclusions. In this Majoron model
the Higgs branching ratio into Majoron is found to be,

\begin{equation}
{\rm BR_{JJ}} = \frac{\Gamma_{JJ}}{\Gamma_{vis}+\Gamma_{JJ}}
\end{equation}where $\Gamma_{vis} = 4.07$~MeV for 
$M_H =125$~GeV and,
\begin{equation}
\Gamma_{JJ} = \frac{\lambda_3^2 v^2}{32 \pi M_H}\left(1-\frac{4M_J^2}{M_H^2}\right) 
\end{equation}

\begin{figure*}[!t]
\centering
\mbox{\includegraphics[width=\columnwidth]{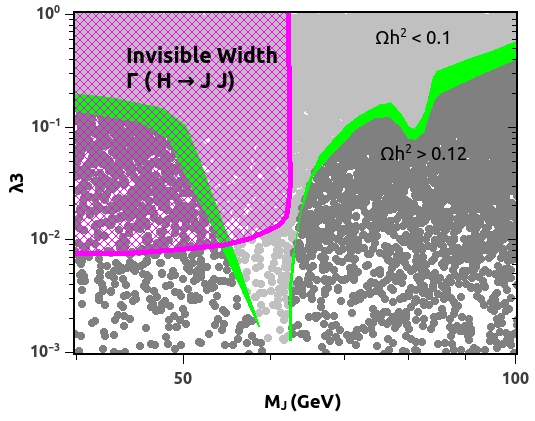}\quad\includegraphics[width=\columnwidth]{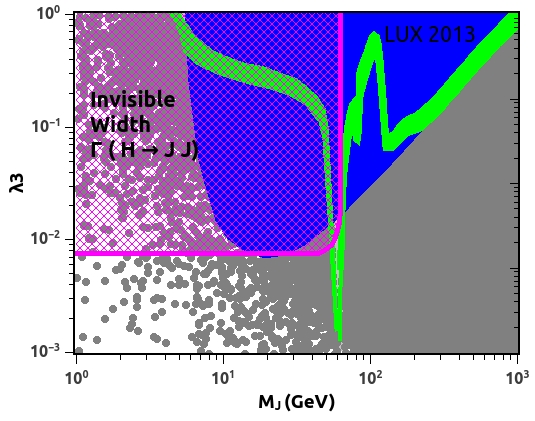}}
\caption{The viable  excluded parameter space of the Majoron dark matter model. The light gray scatter points on the top yield $\Omega h^2 <0.11$, whereas the dark gray ones towards the bottom produces $\Omega h^2 <0.12$. The fine green line between the two gray regions sets  $0.11 <\Omega h^2 <0.12$. The pink shaded region is ruled out by the Higgs invisible decay width taking the face value $BR (H \rightarrow JJ) \leq 5\%$. The blue region is excluded by LUX 2013 data. The parameter $\lambda_3$ is the coupling which connects the dark and visible sector through the Higgs portal. In the left panel we zoomed in the Higgs resonance region and the invisible decay bound, so one can clearly see the latter bounds requires $\lambda_3 < 10^{-2}$. In the right panel we included the LUX constraint in blue. Apart from the resonance region we conclude that the simplest thermal Majoron model studied in the previous section is excluded by direct detection data for $M_J < 225 $~GeV. Moreover, future XENON1T and LUX running are expected to exclude the entire 1GeV-TeV mass range, except the Higgs resonance.}
\label{Graph1}
\end{figure*}

With this in mind we have derived the bound shown in
Fig.\ref{Graph1} (pink shaded region). There we show that $\lambda_3$ has to be smaller than $\sim 10^{-2}$. It is important to point out that there is a lower bound of $\lambda_3 \, > \, 10^{-8}$ for the Majorons to get thermalized down to the electroweak epoch \cite{Burgess:2000yq}.

Regarding the dark matter observables we have computed the relic abundance and the scattering cross section of the Majoron as a function of the two relevant parameters $\lambda_3$ and the Majoron mass. 
The relic density of singlet Majoron
is driven by the s-channel annihilation into SM particles and sub-dominantly determined by the annihilation into hh, through the quartic scalar interaction $J^2 H^2$ in Eq.\ref{jjhh} and by the t-channel Majoron exchange. In Fig.\ref{Graph1} we exhibit the under-abundant $\Omega h^2 < 0.1$ (light gray) and the over-abundant $\Omega h^2 > 0.12$ (dark gray) parameter space, as well as region determined by the green points which yields the right abundance. 
\footnote{There is a small resonance in the Top quark mass. This is due to our numerical calculation using
micrOMEGAS \cite{micromegas}. Micromegas computes tree-level calculations into two-body
final states only, except when the gauge bosons $W,Z$ are in the final state. Therefore micrOMEGAS does not provide a very accurate approximation
close to the threshold for producing gauge boson pairs as they miss the 3- and 4-body final states from virtual decays as well as QCD corrections for quarks in the final state. However, the inclusion of these corrections induce only mild differences in the abundance for the model of interest. The reader can easily see that our results agree well with the results of Ref.\cite{Cline:2013gha,Cline:2012hg} where such processes have been accounted for.
Therefore, our conclusions are unaffected by the insertion of these corrections. See Refs. for similar studies of the Higgs portal \cite{hportalmodels}.}

In the left panel of Fig.\ref{Graph1} we have zoomed in the resonance region in order to clearly show that the Higgs invisible decay width bound requires $\lambda_3 \lesssim 10^{-2}$ assuming a ${\rm BR_{JJ}} \leq 5\%$. If we had assumed a 10\% limit instead the would have found $\lambda_3 \lesssim 3\times 10^{-2}$. Additionally, the green fine line between the light and dark gray regions reproduces the right abundance $0.11 <\Omega h^2 < 0.12$ according to PLANCK data \cite{PLACKcoll}. 

In the right panel of Fig.\ref{Graph1} we display in blue the region ruled out by the LUX bound based on the 2013 data \cite{LUXbound}. One can clearly conclude that a thermally produced singlet Majoron is ruled out by direct detection data for $M_J < 225$~GeV, unless we are sitting at the Higgs resonance. Moreover, future direct detection running coming from XENON1T and LUX are forecasted to exclude the entire GeV-TeV mass range. In order to circumvent this result one might need to evoke the presence of new particles in order to use them as potential co-annihilation channels and consequently suppress the abundance. In other words, pushing down the parameter space which yields the right abundance. Conversely, one could invoke a non-thermal Majoron where the number density is set by a decaying modulus without further annihilation. 
%
%\begin{figure}[!h]
%\centering
%\includegraphics[scale=0.6]{result1.pdf}
%\includegraphics[width=1\columnwidth]{result1.jpeg}
%\caption{The gray scatter points yield $\Omega h^2 <1$. The blue curve is set the parameter space which reproduces the correct abundance $\Omega h^2 =0.1$. The shaded red region is excluded by LUX 2013 bounds. The parameter $\lambda_3$ is the coupling which connects the dark and visible sector through the Higgs portal. One can easily conclude that the simplest thermal Majoron model studied here is excluded by direct detection data.}
%\label{Graph1}
%\end{figure}

In addition to the abundance and direct detection bounds, it is important to also consider indirect detection bounds in this model. Late decay of Majorons to neutrinos would produce too much power at large scales, through the late integrated Sachs-Wolfe effect, thus spoiling the CMB anisotropy spectrum. WMAP third year data can be used to constrain \cite{CMBmajoron},
\begin{equation} \label{Majtoneutrinos}
\Gamma_J < 6.4 \times 10^{-19} s^{-1}
\end{equation}
From Eq.\ref{Eqlifetime} one may notice for the mass range of interest ($<1$~TeV) this limit is easily obeyed as long as a huge lepton number violation scale is assumed. Furthermore, bounds coming from measurements of the atmospheric neutrinos background flux might be important as well \cite{Covi:2009xn,Esmaili:2012us}, but for Majorons lighter than $1$~TeV the LUX bounds are the most relevant because of the the cutoff in the energy flux. See Fig.2 of \cite{Covi:2009xn}.
Because in this section we are studying the simplest see-saw Majoron model, no radiative decay is present and therefore no bounds from x- or $\gamma$-ray apply. However, in the next section we briefly discuss the case of light Majoron dark matter in more general seesaw models, which can possibly explain the recent keV line observed by \cite{KeVline} and show the most current constraints on x- and $\gamma$-rays lines in Fig.\ref{Graph2}.

\section{Light Majorons and the keV Line} \label{kevmajoron}

In this section, we discuss light Majorons with mass $\mathcal{O}($keV$)$ as dark matter candidates. We will assume that the coupling to the Higgs portal is small enough to evade invisible decay width and other bounds described in the previous section.

%Higgs decay to Majoron dark matter is kinematically allowed for light Majorons, and the decay width is $\Gamma_{H \rightarrow JJ} \, \sim \, \frac{1}{8\pi m_H} (\lambda_3 v)^2$. This is now constrained by limits on Higgs invisible decays at the LHC. For $\lambda_3 \lesssim 10^{-2}$, one obtains Br$_{H \rightarrow JJ} \, \sim \, 10 \%$, which is allowed by current CMS and ATLAS data \cite{Ellis:2013lra}. We will assume that $\lambda_3$ is in this allowed range. Clearly, for viable light Majoron dark matter, at least percent-level fine-tuning in the Higgs coupling is necessary. %As an extreme limit, we will consider the cosmologically interesting case of keV-scale Majoron dark matter. 

The relic density of light Majoron dark matter at the present time is determined by its number density relative to photons at the time of decoupling, and should also account for the finite decay width in Eq.\ref{Majtoneutrinos}. %The initial number density could be set by a thermal or non-thermal history. 
In general one might write the relic density as 
\be
\Omega h^2 \, = \, \frac{78}{g_*}\frac{m_J}{1 \,\, {\rm keV}} e^{-t_0/\tau_J}
\ee
%
%$\Omega h^2 \, = \, \beta \frac{m_J}{1.25 \,\, {\rm keV}} e^{-t_0/\tau_J}$, 
for the case of thermal freeze-out at the time of decoupling (and $g_* = 106.75$). Clearly, a Majoron of mass $\sim 0.1$ keV satisfies the relic density constraint via thermal freeze-out. In general, however, higher masses are possible - for example, thermal freeze-in of the Majoron through the sterile neutrino portal can accommodate masses between 1 keV and 3 MeV \cite{Frigerio:2011in}. Various non-thermal mechanisms of Majoron production have also been considered in the literature \cite{Rothstein:1992rh}, which can give a heavier Majoron in the keV range to satisfy the relic density. %The case of $\beta \, < \, 1$ parametrises this ignorance of the early history.   

%For the case of thermal freeze-out, one has $\beta = 1$.  %\cite{CMBmajoron}.
%
In the simplest Majoron model described in the previous sections, there is no coupling to photons. However, a coupling like $J F \tilde{F}$ is natural at loop level in see-saw models with the presence of scalar $\Delta$ that are triplets under $SU(2)_L$ \cite{Berezinsky:1993fm}.
%
% In order to derive more general conclusions we will consider the Majoron model with the presence of a scalar triplet plus a singlet scalar following the procedure described in \cite{CMBmajoron}. 
% 

Following \cite{Berezinsky:1993fm}, one extends the Lagrangian in Eq.~\ref{yukawa} by also adding the term $\lambda^{\prime} \bar{L}^{} \Delta L$, where $\Delta$ has hypercharge 1 and lepton number $-2$ and $\lambda^{\prime}$ is the relevant Yukawa coupling. In a calculation similar to that in the previous sections, one finds that the Majoron now has a non-zero component along the Higgs doublet, and it is given by 
\be
J \, \sim \, \frac{v^2_3}{uv} \phi + \frac{v_3}{u} \Delta + \sigma, 
\ee
where $u, v,$ and $v_3$ are the singlet, doublet, and triplet Higgs vevs respectively, arranged to satisfy the relation 
%$u v_3 \sim v^2$, with 
$u \gg v \gg v_3$. The Majoron is thus still mostly along the singlet direction, although the doublet projection will be vital for us.

The decay rate to neutrinos (which controls the lifetime and cosmological stability of the Majoron) is controlled mainly by the profile of the Majoron along the singlet Higgs direction in accordance with Eq.~\ref{Eqdecaywidth}. This can be seen by writing out the coupling of the Majoron to the mass eigenstate neutrinos explicitly to leading order, most conveniently by utilizing the symmetry properties of the Higgs potential
\be
g_{Jij} \, = \, - \frac{m^{\nu}_{i} \delta_{ij}}{2 u} + \ldots
\ee
The profile along the triplet direction may also lead to interesting physics, such as the possible decay of neutrinos, as elucidated in \cite{Berezinsky:1993fm}. Models with the triplet Higgs $\Delta$, independently of the Majoron, have been widely studied in the context of Higgs physics, LHC searches, and Left-Right models \cite{triplethiggs}.

The leading decay width of the Majoron is to neutrinos. But now there is also a coupling of the Majoron to charged fermions coming from the Higgs projection, whose strength goes like 
\be
g_{J\gamma \gamma} \, \sim \, \frac{v^2_3}{uv} m_f (T_{3f}) \bar{f} \gamma_5 f J \,\,,
\ee 
where $m_f$ and $T_{3f}$ are the mass and weak isospin of the charged fermions. This further induces at loop level a coupling to photons, with a decay width given by
\be
 \Gamma_{J \gamma \gamma} = \frac{\alpha^2}{64 \pi^3} \frac{v^4_3}{v^4 u^2 }\frac{m^3_J}{\Lambda^2} \,\,, 
\ee
where $\Lambda = \frac{1}{\Sigma_f N_f q^2_f (-2T_{3f})\frac{1}{12} \frac{m^2_J}{m^2_f}}$. Note that $q_f$ and $N_f$ are the electric charge and color factor of the SM fermions $f$ that couple to the Majoron \cite{CMBmajoron}. 

\begin{figure}[!h]
\centering
\includegraphics[width=1\columnwidth]{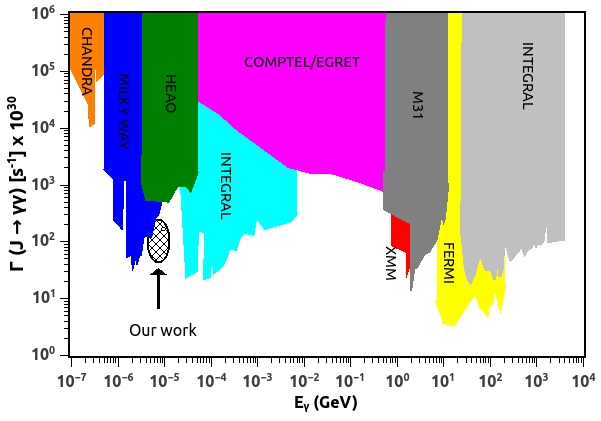}
\caption{We show the predicted line emission in the Majoron model with Higgs triplets. Besides we exhibit the line emission constraints on the
decay rate into two monoenergetic photons from left to right: CHANDRA LETG NGC3127 (orange) \cite{CMBmajoron}, Milky Way (blue)\cite{M31bound}, HEAO observations of the diffuse x-ray background (green) \cite{HEAO}, INTEGRAL diffuse background \cite{INTEGRAL2}, COMPTEL and EGRET Telescopes (pink) \cite{COMPTELEGRET}, M31 (dark gray) \cite{BULLETCluster}, XMM observations of the Milky Way  (red) \cite{XMMbound}, Fermi-LAT search (yellow) \cite{fermi},INTEGRAL SPI line search in the Milky Way halo (light gray) \cite{INTEGRAL}.}
\label{Graph2}
\end{figure}

In the analysis of \cite{KeVline}, the requirements to explain the $3.5$ keV line observed by XMM-Newton observatory in the spectrum of 73 galaxy clusters are 
\bea
m_J  &\sim & 7 \,\,\,\, {\rm keV}, \nonumber \\
\Gamma_{J \gamma \gamma} &\sim & 10^{-28} s^{-1} \,\,.
\eea
The required decay width can be obtained for $v_3 \sim 0.1 - 1$ GeV, with $u \sim 10^4$ TeV %from $u v_3 \sim v^2$. 
The triplet vev in this range is compatible with electroweak precision constraints \cite{Erler:1999ub}. We note that this value of $u$ also satisfies the constraint on the leading order decay to neutrinos, Eq.~\ref{Majtoneutrinos}. Moreover, this result is also consistent x-rays bounds coming from a variety of sources. In Fig.\ref{Graph2} we show the exclusion regions as well as the result predicted in the Majoron model with a Higgs triplet discussed above. 

Since we will be interested in $m_J \sim \mathcal{O}(7)$ keV, the relic density constraint requires a mechanism that is different from thermal freezeout. We mention a few more points about satisfying the relic density in the scenario above. As stated before, thermal freeze-in is an option to satisfy the relic density with a $\sim \mathcal{O}(10)$ keV Majoron. In this scenario, the sterile neutrinos annihilate into the Majoron with a rate that is too low for the Majoron to thermalize. The under-abundant Majoron density reaches a plateau when the temperature $\sim$ mass of the sterile neutrinos. The process thus depends on the annihilation cross section $\sigma(\bar{N}_{R}^{c}N_{R} \rightarrow JJ)$ of the sterile neutrinos into the Majoron, which depends on the mass $m_J$ and the Yukawa coupling $h$ of the Majoron with the sterile neutrinos, as in Eq.~\ref{yukawa}. In fact, for the Majoron mass $\sim 7$ keV needed to satisfy the x-ray signal, we require $h \sim 10^{-3}$ for thermal freeze-in to give the correct relic density \cite{Frigerio:2011in}. %This small Yukawa in fact helps satisfy the constraint on the Majoron lifetime Eq.~\ref{Majtoneutrinos}.

In summary, a Majoron produced by thermal freeze-in or non-thermally might be a potential explanation to this 3.5 keV line recently observed while still being consistent with other searches. Alternative solutions to this keV line have been put forth \cite{KeVmodels}. 

\section{Conclusions}

Majorons are Nambu-Goldstone bosons arising from the spontaneous breaking of lepton number symmetry by a complex scalar. The Majoron mass is model-dependent, although it is expected to be small, due to explicit lepton symmetry breaking by quantum gravity effects. On the other hand, explicit soft terms can be introduced to make the Majoron mass large.

These scalar dark matter models have several interesting features: 

$(1)$ they are examples of decaying dark matter, with the decay being mainly to neutrinos, suppressed by the scale of lepton symmetry breaking. 

$(2)$ in the heavy (GeV - TeV) Majoron case, this model is a UV realization of the effective Higgs-portal scalar dark matter framework. With this in mind we have computed the Higgs invisible decay width into Majorons, abundance, direct and indirect detection observables and compared them with the most current data available. The Majoron has a somewhat large parameter space that can reproduce the right abundance as shown in Fig.1. In case the Majoron is sitting at the Higgs resonance the model is consistent with all current bounds. Otherwise the recent constraint on the scattering cross section reported by LUX in 2013 decisively rules out the thermal Majoron window for $M_J < 225$~GeV in the simplest Majoron model which has a singlet scalar and heavy-right handed neutrinos. Additionally, future direct detection running coming from XENON1T and LUX are expected to rule out the entire GeV-TeV mass range. Therefore either non-thermal production mechanisms or the inclusion of new particle to exploit co-annihilation channels are required to circumvent this conclusion. Another alternative would be the inclusion of new particles to play to role of the mediator in new annihilations channels. 

We have also discussed indirect detections bounds. In particular the late decay of Majorons may distort the CMB power spectrum, therefore a bound of $\Gamma < 10^{-19} s^{-1}$ is required. This enforces the Majoron mass to not be much greater than the TeV scale for a large scale where the lepton number is softly broken. If one pushes down the latter scale the upper bound on Majoron is rapidly strengthened according to Eq.16.

$(3)$ lastly, we discussed a Majoron model which has a Higgs triplet in its spectrum in light of the recent 3.5~keV line and concluded that the required signal could be obtained for a lepton number violation scale $\sim \mathcal{O}(10^4)$ TeV. We noted that the thermal freeze-out scenario does not address such a line because the mass of the Majoron cannot be larger than $\sim 0.1$~keV. However, if one uses thermal freeze-in through the sterile neutrino portal to set the Majoron relic density (alternatively, a non-thermal mechanism may also work), then it is possible to have the Majoron as the possible candidate to this $3.5$~keV signal while being consistent with other x-ray searches according to Fig.2. We provided a proof of concept that the observed keV line can be obtained in this class of models, for plausible choices of the model parameters. The relation between light decaying dark matter, line emission bounds, and thermal freeze-in certainly warrants further study \cite{farinaldokuver}.

\section*{Acknowledgement}
FQ is partly supported by US Department of
Energy Award SC0010107 and the Brazilian National Counsel for Technological and Scientific Development (CNPq). KS is supported by NASA Astrophysics Theory Grant NNH12ZDA001N.
The authors would like to thank Will Shepherd and Patrick Draper for useful discussions.

\end{document}